%% file: springer.tex
\documentclass[runningheads]{llncs}
\usepackage{amssymb,amsfonts}
\usepackage{graphicx}
\usepackage{cite}
\usepackage{algorithmic}
\usepackage{graphicx}
\usepackage{textcomp}
\usepackage{xcolor}
\usepackage{xspace}
\usepackage{hyperref}
\usepackage{tikz}
\usepackage{tikz-uml}
\usepackage{courier}
\usepackage{listings}
\usepackage[utf8]{inputenc}
\usepackage[T1]{fontenc}
\usepackage{silence}
\usepackage{subcaption}

\newcommand{\dtn}{DTN7\xspace}

\newcommand{\footurl}[1]{\footnote{\url{#1}}}

\begin{document}

\title{\dtn: An Open-Source Disruption-tolerant Networking Implementation of Bundle Protocol 7}
\author{
	Alvar Penning\inst{1} \and
	Lars Baumgärtner\inst{3} \and 
    Jonas Höchst\inst{1,2} \and \\
	Artur Sterz\inst{1,2} \and 
    Mira Mezini\inst{3} \and
	Bernd Freisleben\inst{1,2}
}

\authorrunning{Penning et al.}
\titlerunning{\dtn: An Open-Source Implementation of Bundle Protocol 7}

\institute{
    \textit{Dept. of Math. \& Computer Science, Philipps-Universität Marburg, Germany} \\
	\email{\{penning, hoechst, sterz, freisleb\}@informatik.uni-marburg.de} \\ \and
    \textit{Dept. of Electr. Engineering \& Information Technology, TU Darmstadt, Germany} \\
	\email{\{jonas.hoechst, artur.sterz\}@maki.tu-darmstadt.de} \and
	\textit{Dept. of Computer Science, TU Darmstadt, Germany} \\
	\email{\{baumgaertner, mezini\}@cs.tu-darmstadt.de}
}

\maketitle

\begin{abstract}
\input{00_abstract}

\end{abstract}

\input{01_intro}
\input{02_relwork}
\input{03_bp7}
\input{04_implementation}

\input{05_evaluation}
\input{06_conclusion}

\section*{Acknowledgement}
This work is funded by the HMWK (LOEWE Natur 4.0 and LOEWE emergenCITY) and the DFG (SFB 1053 - MAKI).


\bibliographystyle{splncs04}
\bibliography{literature}

\end{document}

%% file: 00_abstract.tex
In disruption-tolerant networking (DTN), data is transmitted in a store-carry-forward fashion from network node to network node. 
In this paper, we present an open source DTN implementation, called \dtn, of the recently released Bundle Protocol Version 7 (draft version 13).
\dtn is written in Go and provides features like memory safety and concurrent execution.
With its modular design and interchangeable components, \dtn facilitates DTN research and application development.
Furthermore, we present results of a comparative experimental evaluation of \dtn and other DTN systems including Serval, IBR-DTN, and Forban. 
Our results indicate that \dtn is a flexible and efficient open-source multi-platform implementation of the most recent Bundle Protocol Version 7.

%% file: 01_intro.tex
\section{Introduction}
\label{sec:intro}

Delay- or disruption-tolerant networking (DTN) is useful in situations where a reliable connection to a communication infrastructure cannot be established, e.g., during environmental monitoring in remote areas, if telecommunication networks are destroyed as a result of natural or man-made disasters, or 
if access is blocked due to political censorship. In DTN, messages are transmitted hop-to-hop from network node to network node in a store-carry-forward manner. 
There might be larger time windows between two transmissions, and the next node to carry a message might be reached opportunistically or through scheduled contacts.

There are several mobile DTN appications, such as FireChat~\cite{garden2015firechat} and Serval~\cite{gardner2011serval}, that rely on peer-to-peer networks of smartphones, where
the pre-installed Wi-Fi or Bluetooth hardware of the mobile devices is used to create a large mesh network.
{\textmu}PCN~\cite{feldmann2015upcn} is a special purpose DTN application for planetary communication, and IBR-DTN~\cite{doering2008ibr} is a popular DTN platform, but does not implement the recently released Bundle Protocol (BP) Version 7~\cite{dtn_bp7v13}.

In this paper, we present \dtn, which (to the best of our knowledge) is the first and only freely available, open source implementation of the most recent draft of Bundle Protocol Version 7 (BP7) (draft version 13).
\dtn is designed to offer extensibility by allowing developers to easily replace or add individual components. 
\dtn is a general purpose DTN software with support for several use cases, such as enabling communication in disaster scenarios or providing connectivity in rural areas. 
Our contributions are:
\begin{itemize}
    \item We provide a memory-safe and concurrent open-source implementation of BP7 (draft version 13), written in the Go programming language.
    \item With its highly modular design and its focus on extensibility by providing interfaces to all important components, \dtn is a flexible basis for DTN research and application development for a wide range of scenarios.
    \item We compare \dtn with other well-known DTN systems including Serval, IBR-DTN, and Forban, using the CORE network emulation framework.
    \item Several experiments to mimic different DTN test cases, i.e., a chain of up to 64 nodes with different payload sizes,
    are conducted.
    \item The presented \dtn 
    software\footnote{\url{https://github.com/dtn7/dtn7-go}},
    the evaluation framework and its 
    configurations\footnote{\url{https://github.com/dtn7/adhocnow2019-evaluation}},
    and the experimental 
    fragments\footnote{\url{https://ds.mathematik.uni-marburg.de/dtn7/adhoc-now_2019.tar.gz}}
    are freely available.
\end{itemize}

The paper is structured as follows. Section~\ref{sec:relwork} discusses related work. 
In Section~\ref{sec:bp7}, we briefly explain BP7. Section~\ref{sec:implementation} discusses \dtn's design and implementation.
Section~\ref{sec:evaluation} describes experimental results. Section~\ref{sec:conclusion} concludes the paper and outlines areas of future work.

%% file: 02_relwork.tex
\section{Related Work}
\label{sec:relwork}

This section briefly reviews relevant publications in the area of DTN software.

\subsection{DTN Software Implementations}

IBR-DTN~\cite{doering2008ibr} is a lightweight, modular DTN software for terrestrial use.  
The Interplanetary Overlay Network (ION)
focuses on the aspects of extreme distances in space~\cite{burleigh2007interplanetary}.
DTN2 is the reference implementation of the BP, developed by the IETF DTN working group~\cite{demmer2004implementing}.
These three implementations are based on RFC 5050, i.e., BP Version 6~\cite{rfc5050}.

Designed for small satellites in low earth orbit, {\textmu}PCN can be used to connect different regions of the world.
It also implements BP Version 6, as well as an older draft of version 7~\cite{feldmann2015upcn}.
Furthermore, an older version of BP7 is implemented in Terra~\cite{rightmesh2019Terra}.

Serval focuses on node mobility by providing implementations that run on smartphones, as well as by incorporating different radio link technologies~\cite{gardner2011serval}.
Forban is a peer-to-peer file sharing application that uses common Internet protocols like IP and HTTP to transmit files in a delay-tolerant manner~\cite{dulauny2019forban}.
With FireChat~\cite{garden2015firechat}, it is possible to send messages via DTN without relying on Internet access or direct peer contacts.

Many of the mentioned DTN systems implement the BP as specified in RFC 5050~\cite{rfc5050}.
While some implement a draft of BP7, none of them implements the most recent draft.
Serval, Forban, and FireChat have their own protocol definitions, which are not compatible with the BP.
Furthermore, the mentioned implementations cannot be extended in a modular manner, are not written in developer-friendly high-level programming languages and are not intended as general purpose DTN platforms, but are designed for specific use cases.
FireChat is not freely available, and thus cannot be extended.

\subsection{DTN Software Evaluations}

IBR-DTN, DTN2, and ION were evaluated by Pottner et. al~\cite{pottner2011performance}. For a payload of 1 MB, DTN2 and IBR-DTN produced almost identical results. ION was slower in the conducted measurements. Furthermore, the interaction of the three DTN implementations was evaluated by transferring bundles between them, and the times measured varied significantly.

IBR-DTN was used to evaluate the connection between a stationary DTN node and a moving vehicle~\cite{doering2008ibr}. This vehicle passed the stationary node at an average speed of 20 km/h, and the transmission rate was measured in relation to the distance. Data could be transmitted within a range of about 200 meters.

Serval was experimentally evaluated in our previous work~\cite{baumgaertner2016serval}, for scenarios with 48 nodes in a hub topology, 64 nodes in a chain topology, and 100 nodes in disjoint islands connected over time. The results indicate that Serval can achieve high network loads, while CPU usage remains relatively low.

%% file: 03_bp7.tex
\section{Bundle Protocol Version 7}
\label{sec:bp7}

This section gives an overview of bundle protocols, referring to RFC 4838~\cite{rfc4838} and the current version of the Bundle Protocol (BP) ~\cite{dtn_bp7v13}. The latter has version number 7 and is currently still in active development. We discuss the status of the 13th draft from April 2019 below.

\begin{figure}[tbph]
    \centering
    \includegraphics[width=.90\columnwidth]{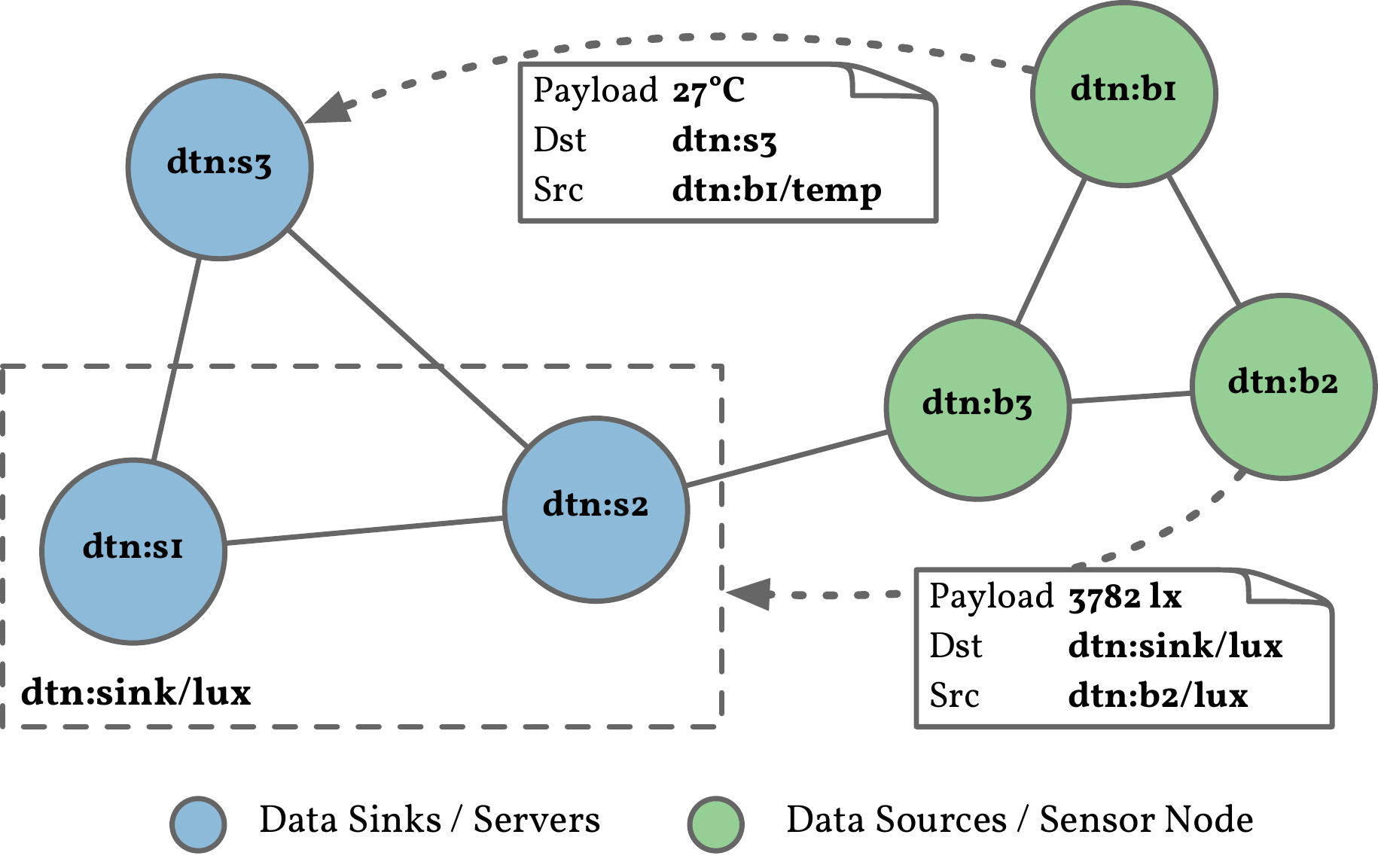}
    \caption{Example sensor node scenario with multiple endpoints.}
    \label{fig:nodes_endpoints}
\end{figure}

\subsection{Basic Concepts}

\subsubsection{Endpoints.}
\label{ssec:dtn-actors}

In DTN, there are nodes and endpoints.
Nodes exchange bundles according to the store-carry-forward principle. 
Bundles are addressed at endpoints, or more precisely, their characterizing \textit{Endpoint Identifier} (EID), which might not be a currently existing part of the network.
Fig.~\ref{fig:nodes_endpoints} shows an example of a scenario, where sensor nodes produce readings to be consumed by data sinks.
The temperature bundle is addressed directly to \texttt{dtn:s3}, where the lux bundle is headed to \texttt{dtn:sink/lux}, an EID that is handled by two nodes, and thus a multicast. BP7 is endpoint scheme agnostic and supports the null endpoint for anonymous bundles.
In BP version 6, only endpoints are defined, so it is not possible to address dedicated nodes.

\subsubsection{Bundles and Blocks.}

Packets in a DTN consist of multiple \textit{Blocks} to form logical units called \textit{Bundles}.
In Fig.~\ref{fig:bundle-sketch}, an example bundle containing the mandatory Primary Block, and two Canonical Blocks, namely a Hop Count Block and the actual Payload Block, is shown, following the example of Fig.~\ref{fig:nodes_endpoints}.

\begin{figure}[tbph]
    \centering
    \resizebox{.9\textwidth}{!}{
        \begin{tikzpicture}
            \begin{umlpackage}[fill=white]{Bundle}
                \umlclass[rectangle split parts=2, fill=white]{Primary Block}{
                    Version: \texttt{7} \\
                    Control Flags: \\
                      \quad \textit{Status requested for reception} \\
                    CRC Type: \textit{None} \\
                    Destination EID: \texttt{dtn:sink/lux} \\
                    Source node EID: \texttt{dtn:b2} \\
                    Report-to EID: \texttt{dtn:b2} \\
                    Creation Timestamp: (\texttt{0}, \texttt{23}) \\
                    Lifetime: \texttt{3600000}
                }{}
                \umlclass[rectangle split parts=2, fill=white, x=4.3, y=0.775]{Hop Count Block}{
                    Type Code: \texttt{9} \\
                    Number: \texttt{2} \\
                    Control Flags: \textit{None} \\
                    CRC Type: \textit{None} \\
                    Data: (\texttt{64}, \texttt{42})
                }{}
                 \umlclass[rectangle split parts=2, fill=white, x=7.65, y=0.775]{Payload Block}{
                    Type Code: \texttt{1} \\
                    Number: \texttt{1} \\
                    Control Flags: \textit{None} \\
                    CRC Type: \textit{None} \\
                    Data: \texttt{0E C6}
                }{}
            \end{umlpackage}
         \end{tikzpicture}
    }
    \caption{A bundle transmitting a lux value from \texttt{dtn:b2} to \texttt{dtn:sink/lux}.}
    \label{fig:bundle-sketch}
\end{figure}

\paragraph{Primary Block.}

Each bundle begins with a (since BP7 immutable) \textit{Primary Block} (see  Fig.~\ref{fig:bundle-sketch}),  containing 
meta-information about the bundle with the following fields:
Version;
Bundle Processing Control Flags to provide information on the bundle, including fragmenting and reporting information;
an optional CRC Checksum (added in BP7 and not available in BP version 6);
Destination EID, Source Node ID and Report-To EID, as  endpoints for administrative records regarding this bundle;
Creation Timestamp, consisting of the actual timestamp and an incrementing sequence number;
Maximum Lifetime of a bundle, expressed in microseconds after creation time;
Fragment Offset and Total Data Length, if fragmented and indicated by the bundle process control flags.

\paragraph{Canonical Block.}
Payload and Extension Blocks in Fig.~\ref{fig:bundle-sketch} are summarized as \textit{Canonical Blocks}. 
These contain a payload in addition to a few block-specific characteristics.
A Canonical Block consists of a Type Code to identify the kind of block, Number to address the specific block, Control Flags and Data. 

The actual payload of the bundle is located in the Payload Block at the end of each bundle.
In addition to sending user data from application programs, status information is also sent within bundles, called Administrative Records, automatically created and sent by DTN software as a response to a previous bundle. 
Extension Blocks are Canonical Blocks containing further information relevant for a DTN router depending on its configuration.
In contrast to BP version 6, the BP7 specification defines the Previous Node Block, Bundle Age Block, and Hop Count Block, and allows user-defined blocks to be added.

\subsection{Node Components}

\subsubsection{Bundle Protocol Agent.}
The \textit{Bundle Protocol Agent} (BPA) offers BP and DTN specific services.
It executes procedures of the BP.
For example, communication between Application Agent and Convergence Layer Adapter (see below) is managed.
The BPA also constructs bundles for the Application Agent.

\subsubsection{Application Agent.}
The interface between the BPA and an application is defined as an \textit{Application Agent} (AA).
A generic AA needs the ability to receive incoming bundles and compose outbound bundles for user applications and services.
Furthermore, an EID must be assigned for local bundle delivery.

\subsubsection{Convergence Layers.}
Bundles are exchanged over connections between nodes of different types and characteristics, and connections are unidirectional or bidirectional, or vary in transmission speed and bandwidth.
Depending on the connection technology used, more or less complex protocols are required for delivery, called \textit{Convergence Layer (CL) Protocols} (CLP).
A \textit{Convergence Layer Adapter} (CLA) is an implementation of a CLP.
There are two CLPs defined by the IETF DTN group to exchange bundles over a TCP connection, the bidirectional TCP Convergence Layer Protocol (TCPCL)~\cite{dtn_tcp} and the unidirectional Minimal TCP Convergence Layer Protocol (MTCP)~\cite{dtn_mtcp}.
In addition to transport layer CLs, there are approaches based on other technologies, e.g., DTN2 defining a Bluetooth and a serial CL, or IRB-DTN featuring an e-mail CL.

%% file: 04_implementation.tex
\section{\dtn}
\label{sec:implementation}

In this section, we present the design and implementation of \dtn. 

\subsection{Requirements Analysis}
There are several requirements that should be satisfied by DTN software. First, DTN software operating on a variety of laptops, smartphones, and routers should run on several hardware architectures (e.g., x86, ARM, and MIPS), based on the most popular operating systems (e.g., Linux, macOS, and Windows).
Second, the individual components of the DTN software should be exchangeable.
For example, there is the need to support different storage backends, CLAs, and DTN routing protocols.
A suitable programming interface enabling concurrent execution is required for the interaction of components.  
Furthermore, a CLA implementation is required as well as a peer discovery mechanism to enable automatic establishment of connections between nodes.
Finally, applications should to be independent of the DTN software, to allow easy creation of further applications and tools.
Thus, a convenient interface between the DTN software and applications is required.

\subsection{Implementation Decisions}

As a result of these requirements, we selected the Go programming language\footnote{\url{https://golang.org}} to develop \dtn.
Go offers a large standard library and is rather developer-friendly.  Its strengths are the simple creation and integration of programming libraries. 
Moreover, Go enforces good style guides and clean code plus provides memory-safety guarantees to increase security and stability of written programs. Thus, Go makes maintaining code and bringing in new developers very easy.
The source code including all required dependencies are compiled into a single, static executable, removing the need for interpreters or further libraries.
Furthermore, the Go compiler allows simple (cross-)compilation for many operating systems and processor architectures.
The concept of concurrency is implemented in Go through the interaction of Goroutines and Channels; concurrency was one of the design priorities of the language designers. 

To support exchangeability of \dtn's components, we structured our implementation into \textit{Bundles} and its corresponding \textit{Store}, \textit{Convergence Layer Adapters}, \textit{Peer Discovery}, the \textit{Application Agent}, \textit{Routing}, and the \textit{Core} package needed to connect the individual packages.
The modules in the these packages are designed as generic interfaces and example implementations, e.g., there exists an interface for routing in general and an epidemic routing implementation.
We decided to use MTCP for exchanging messages between two \dtn nodes due to its simplicity. 
A third party application can also use parts of \dtn as a library to, e.g., create and serialize bundles via the corresponding package.
To make programming of applications against these interfaces simple and programming language independent, we decided to use a RESTful API.

\subsection{\dtn Architecture}
\begin{figure}[t]
    \centering
    \includegraphics[width=0.9\columnwidth]{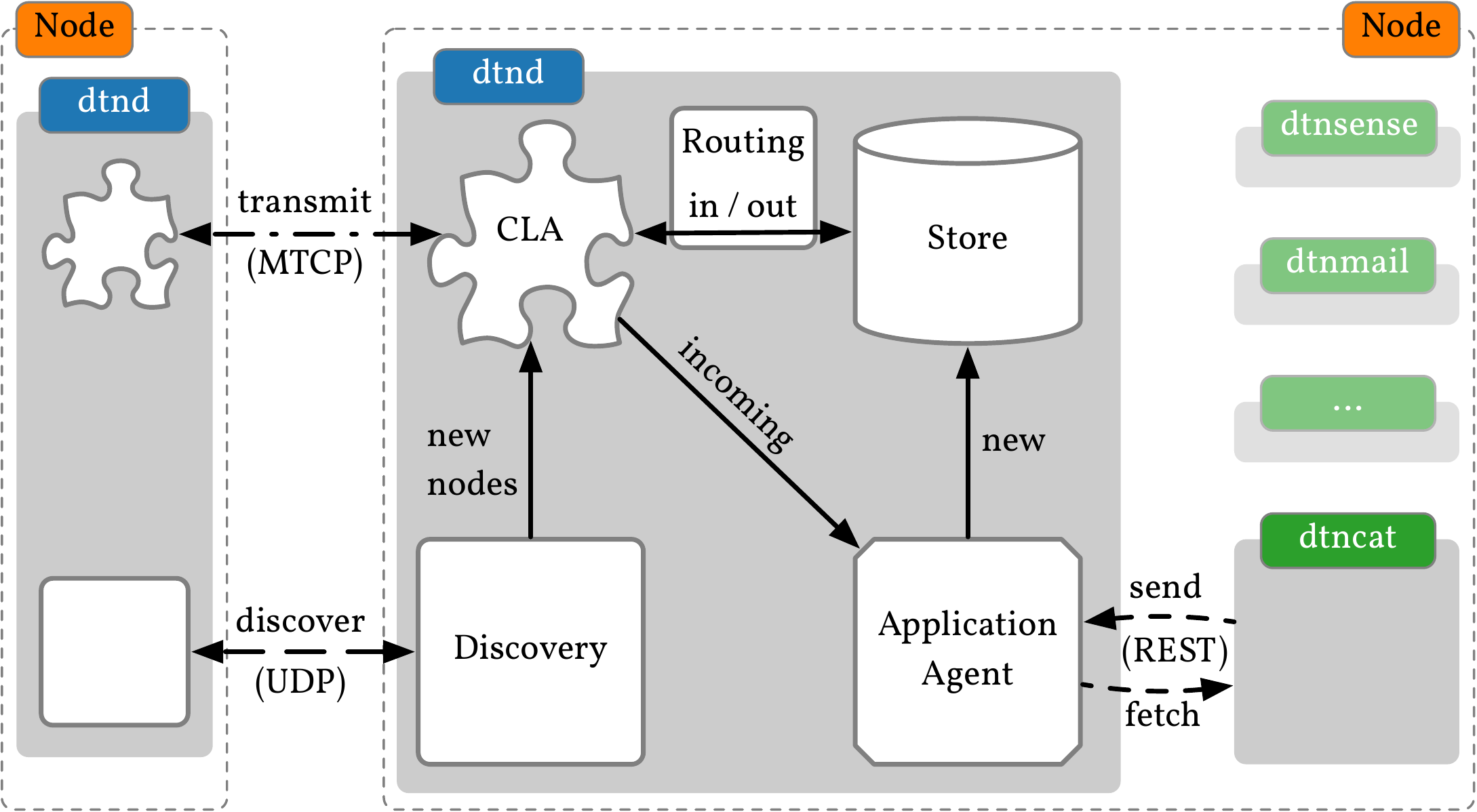}
    \caption{Architecture and data flow in \dtn}
    \label{fig:overview}
\end{figure}

Fig.~\ref{fig:overview} shows the modules of \dtn and their interaction.
The arrows indicate the way a bundle is internally processed in \dtn. 
The links between two distinct \dtn nodes are shown by both an active CLA and the Discovery on the figure's left hand side. 
Multiple client connections to the AA from within the node are delineated on the node's right hand side.

To store bundles locally, a serialized version as defined in BP7 is written to the file system. 
A central index of all known bundles manages their meta-data and links point to information of the specific file.
This index supports a fast lookup of bundles. The module providing this functionality is called \textit{Store}.

In \dtn, an AA is implemented as a RESTful Web API to support both dispatching and fetching of bundles. 
The API does not interact with entire bundles, but only with a subset of its fields.
This allows a client to send a new bundle by only supplying the destination EID and a payload. 
Such a request can easily be created from the command line or possible third-party software.
When fetched over the API, selected fields of those bundles are returned and the bundles will be removed from the store afterwards.

The concept of different CLs and their CLAs is also present in \dtn's architecture with an implementation of MTCP.
Based on a specific CL's characteristics, bundles might be transferred in a uni- or bidirectional way. Thus, a CLA in \dtn must supply one or multiple modules for inbound and outbound bundle processing.
The unidirectional MTCP is designed using modules for sending and receiving bundles.

To support connections in dynamic networks, a \textit{Peer Discovery} mechanism is provided. It announces a node's existence and listens for potential neighbors.
This discovery mechanism broadcasts all of the node's CLAs continuously and notifies about received CLAs.

The previously defined components are linked together within \dtn's \textit{Core} package. 
A central processing pipeline consumes both newly created and inbound bundles. 
Within this pipeline, a bundle will be marked to be delivered to a subset of known CLAs, to a local AA or to be discarded for later processing or even removed.
The Core's internal links, visualized in Fig.~\ref{fig:overview}, are related to the concept of a BPA, and serve as an interface between CLAs and the AA.

Every bundle that is not addressed to a particular node will be forwarded over one or multiple CLAs to neighboring nodes.
The decision about which CLAs to select is made by a routing algorithm.
To support the use of different routing algorithms, a generic interface needs to be informed about inbound bundles and, furthermore, a tight cohesiveness to the core is required.
\dtn implements an epidemic routing module, which is notified about received bundles, to memorize both sender and receiver.
Before dispatching, the epidemic routing algorithm compiles a subset of known connections which have not received this bundle yet.

Finally, \dtn is also intended to be used as a library and allows fast development of DTN applications. In particular, bundle package creation, serialization, and deserialization  might be useful in other software.

\subsection{Resulting Programs}

\dtn contains a DTN daemon, referred to as \texttt{dtnd} in Fig.~\ref{fig:overview}, for storing and exchanging bundles and interfacing with applications. Currently, an example DTN application (\texttt{dtncat} in Fig.~\ref{fig:overview}) for sending and receiving bundles, implemented as a command line tool, is included. 
\texttt{dtnd} initializes the previously defined modules according to the configuration provided by the user.
\texttt{dtncat} processes user input, which is handed over to \texttt{dtnd}'s AA RESTful interface.
The input is then encapsulated inside the Payload Block of a newly created bundle by \texttt{dtnd}. This bundle's Primary Block will be populated with basic defaults, like disabled CRC, and a delivery report request.
As shown in Listing~\ref{lst:dtncat}, \texttt{dtncat} is called by passing parameters on the command line. 
The first option selects between receiving or sending bundles.
The local \texttt{dtnd}, running the RESTful API, is addressed by the second parameter. 
When sending new bundles, the content is read from the standard input.

\begin{lstlisting}[caption={\texttt{dtncat} example}, captionpos={b}, label={lst:dtncat}]
# Sending a bundle
$ dtncat send http://localhost:8080 dtn:s2 <<< "3782 lx"

# Retrieving a received bundle 
$ dtncat fetch http://localhost:8080 
\end{lstlisting}

%% file: 05_evaluation.tex
\section{Experimental Evaluation}
\label{sec:evaluation}

In this section, we experimentally evaluate \dtn and compare it with other DTN software.

\subsection{Emulation Environment}
To evaluate \dtn in a realistic manner, we emulated up to 64 nodes in the network emulation framework \emph{Common Open Research Emulator} (CORE)~\cite{ahrenholz2010comparison}.
CORE can emulate nodes using Linux namespaces to allow the execution of native binary programs, which is not possible with purely simulation-based approaches like NS-3~\cite{riley2010ns,schwerdel2011tomato}.
All experiments were performed on Intel Xeon E5-2698 CPUs with 80 cores at 2.20 GHz and 256 GB RAM.
To execute the total number of 1,440 experiment runs, we used MACI, a framework for extensive and reproducible experiments~\cite{froemmgen2018maci}.

\subsubsection{DTN Software.}
We compared \dtn with three popular DTN software solutions.
\textit{Serval}\footurl{https://github.com/servalproject/serval-dna/tree/batphone-release-0.93} is a software suite centered around protocols designed for infra\-structure independent communication~\cite{gardner2011serval}.
To be able to transfer files in intermittently connected networks, Serval relies on Rhizome, a custom DTN bundle protocol with epidemic routing.
In our evaluation, we used the latest stable Serval release, which is from April 2016, since the recent development version has stability issues.
\textit{IBR-DTN}\footurl{https://github.com/ibrdtn/ibrdtn} is an implementation of BP Version 6, aimed to be lightweight and fast~\cite{doering2008ibr}.
For comparability, we use the epidemic routing extension instead of the default PRoPHET protocol used by IBR-DTN.
We use the current HEAD of the git repository to include the latest bug fixes.
\textit{Forban}\footurl{https://github.com/adulau/Forban} is mainly used as a local peer-to-peer file sharing application using an epidemic routing protocol based on HTTP.
We used the latest HEAD of the git repository, but had to introduce our own patches to make Forban usable.


\subsubsection{Payload Sizes.}
DTN software is used in multiple applications and scenarios.
Serval, e.g., offers the SMS-like application MeshMS for short text messages.
IBR-DTN can be used in environmental monitoring, where transmission of short audio recordings or images might be required. 
Therefore, we selected four different file sizes, representing a wide range of possible applications.
All files were generated randomly with the same seed for reproducibility in six sizes:

\begin{itemize}
    \item 
\textit{64 KiB} for compressed images or map data;
\item
\textit{1 MiB} representing small images or short audio recordings;
\item
\textit{5 MiB}, e.g., smartphone images and audio recordings;
\item
\textit{25 MiB} representing longer audio recordings or short videos;
\item
\textit{50 MiB} for HD videos typically recorded by smartphones;
\item
\textit{100 MiB}, e.g., 4k smartphone videos \cite{trono2015dtn,schildt2011ibr,baumgaertner2016serval}.
\end{itemize}

\subsubsection{Network Topologies.}
We used a chain topology of three different lengths, where nodes are connected pairwise, to benchmark the different DTN software systems.
The first node is sending a bundle destinated to the last node in the chain.
To get the baseline performance of the interacting components, a chain of two nodes was used.
We measured the time it takes to read the data, serialize the bundle, send it over the network, deserialize it at the receiver and deliver it to the application.
With 32 nodes, the forwarding capabilities were investigated.
For an even larger scenario, we used 64 nodes, to evaluate how the DTN software systems behave when node numbers increase.
We used a bandwidth of 54 MBit/s to match the speed of an IEEE 802.11g network. 


\subsubsection{Measurements.}
To measure CPU utilization for each process on every node, we used \textit{pidstat}, which is part of the \textit{sysstat} package\footnote{http://sebastien.godard.pagesperso-orange.fr/man\_pidstat.html}.
Additionally, \textit{bwm-ng}\footnote{https://github.com/vgropp/bwm-ng} was used for network statistics per node and network interface.
Finally, every used DTN software logged both the timestamp of sending and receiving bundles, such that a detailed analysis of transmission time and network distribution can be performed.

\subsection{Results}

\subsubsection{Transmission Times.}
\label{subsub:transmission_time}

\begin{figure}[t]
    \centering
        \includegraphics[width=0.8\columnwidth]{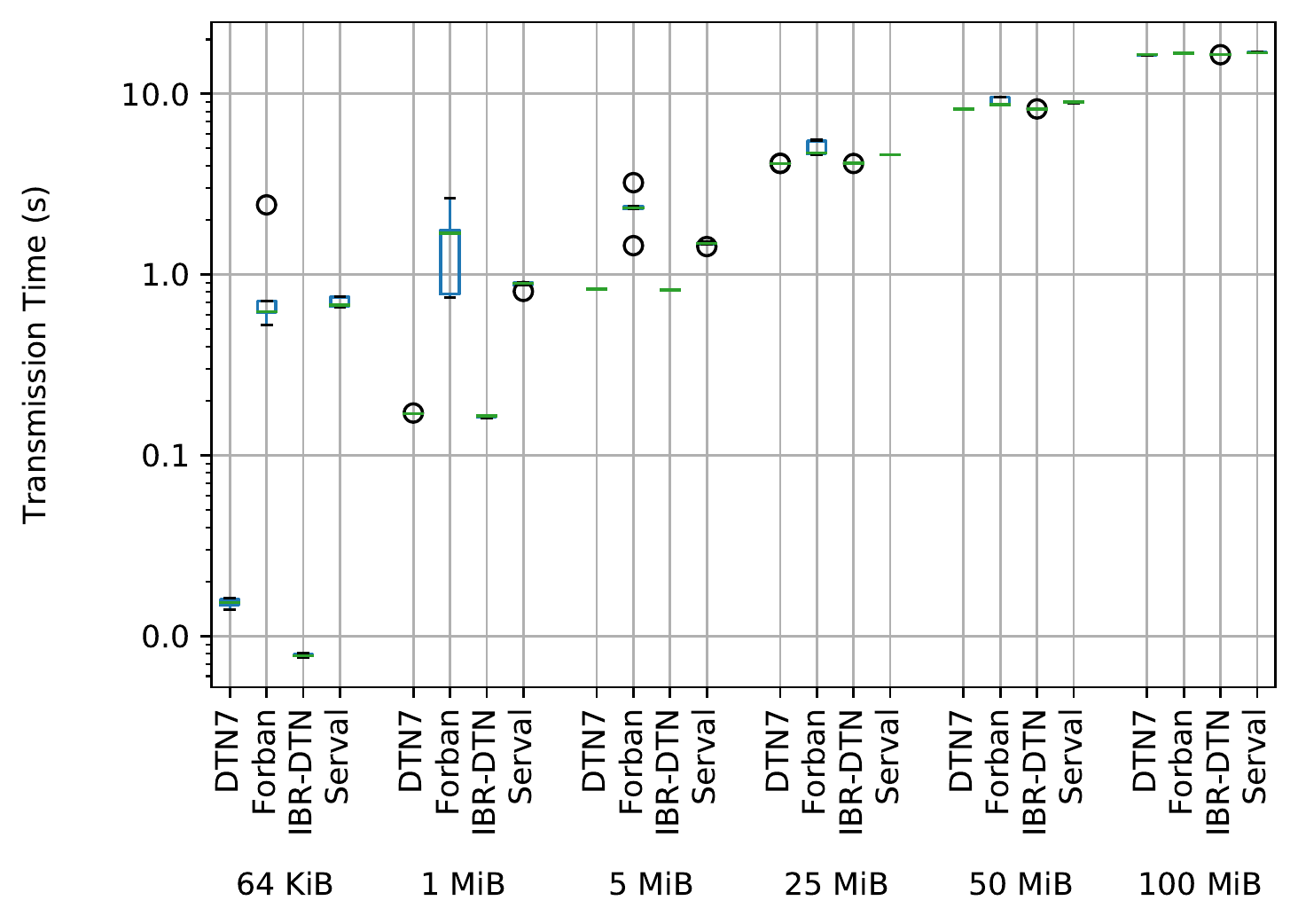}
        \caption{Bundle transmission time for the 1-hop topology and different payload sizes}
        \label{fig:transmissions1}
\end{figure}
\begin{figure}[h]
        \centering
        \includegraphics[width=0.8\columnwidth]{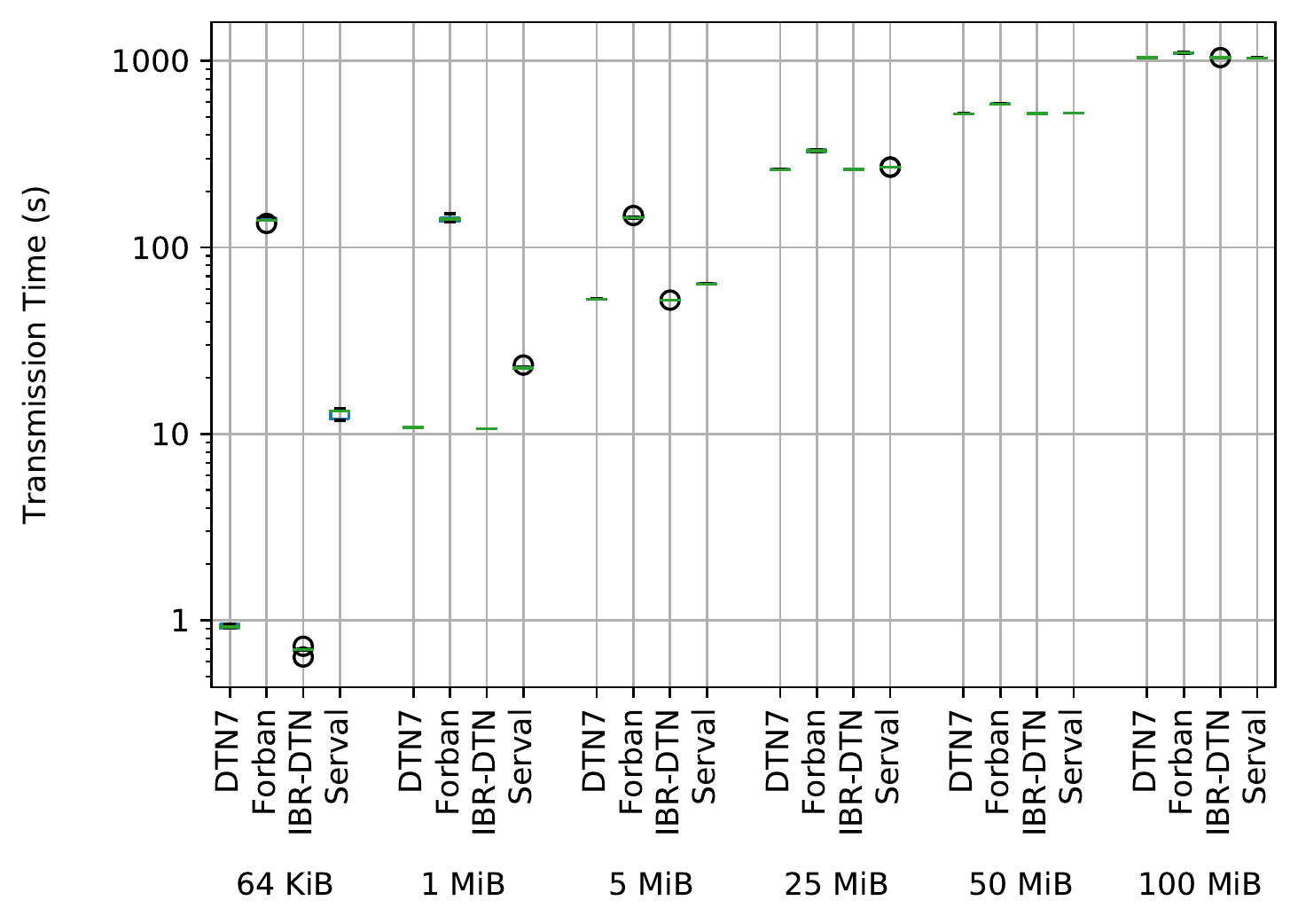}
    \caption{Bundle transmission time for the 64-hops topology and different payload sizes.}
    \label{fig:transmissions64}
\end{figure}

Figs.~\ref{fig:transmissions1} and \ref{fig:transmissions64} show the bundle transmission times on the y-axes and payload sizes on the x-axes for the 1-hop and 64-hops topologies, respectively.
Regardless of chain length and file size, \dtn and IBR-DTN are always the fastest DTN software systems.
The larger the files become, the transfer times of all DTN systems converge.
This is due to the network configuration.
All DTN systems manage to completely fill the 54 Mbit/s available, which is easier to achieve with larger files.
As a result, the transfer times for large files hardly vary at all.

For a single hop, Forban and Serval take about the same time for transmitting files (e.g., about 0.6 seconds for 64 KiB files), but Forban shows a higher variance.
For longer chains and files below 50 MiB, the differences between Forban and Serval are more noticable.
\dtn, however, is still up to 140 times (64 KiB over 1 hop) faster than Serval. 
Particularly in chat or text based applications, the speed advantage of \dtn can be crucial if a message arrives below 0.01 seconds rather than after one second.

These results indicate that both BP6 and BP7 have a relatively small protocol overhead compared to the protocols used by Serval and Forban, which is especially noticeable for small files.
The larger the files or the longer the chain, the less weight the low protocol overhead carries.
Furthermore, it is also remarkable that \dtn, which is written in Go, does not take longer to transmit larger files from end to end in the chain, although IBR-DTN is implemented in C++ and optimized for speed.
In terms of transmission speeds, Forban takes longer than the other DTN software systems, although differences get smaller the bigger the files are. One explanation is that Forban has a pull-based approach where it actively downloads new bundles after an announcement was received. Therefore, the announcement interval is a natural barrier. If quick data exchange is necessary, the other solutions provide better performance.

\begin{figure}[t]
\hspace{-15mm}
    \includegraphics[width=1.2\columnwidth]{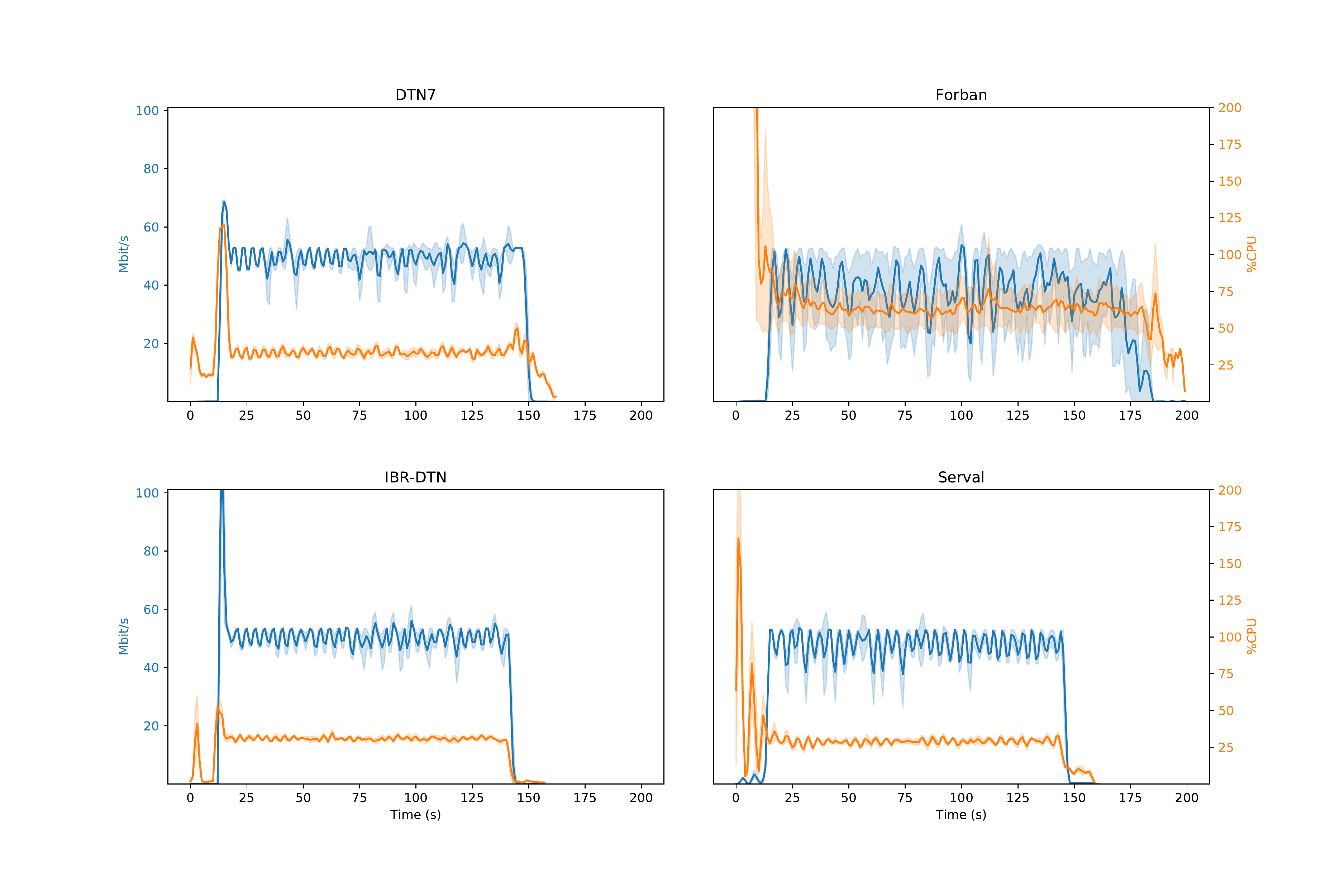}
    \caption{CPU and network usage for transmitting 25 MiB over 32 hops.}
    \label{fig:cpu_net}
\end{figure}

\subsubsection{CPU Usage and Network Utilization.}

Fig.~\ref{fig:cpu_net} shows CPU usage and network utilization for transmitting 25 MiB over 32 hops.
On the x-axes, the time for the entire experiment in seconds is shown, the left y-axes denote the network usage in Mbit/s and the right y-axes show the CPU usage in \%, both of the entire network.
The bold graphs denote the sum over all nodes, averaged over all experiment repetitions.
The shaded areas denote the error band.

\dtn requires about 34.3\% of the available CPU (standard deviation of 16.7\%).
At the beginning of an experiment, \dtn shows a short peak in CPU usage resulting from the first node, where the file is converted to base64, sent to the \dtn AA, which decodes the file again, packs it into a bundle, and starts the transmission.
Further nodes only have to retransmit the bundle and do not require the steps mentioned above.
Forban uses about 163.1\% CPU (646.3\%).
Forban shows a small peak at the start of the experiment, indicating the overhead when starting its daemons, where four Python interpreters have to be started.
Additionally, the file has to be hashed at the beginning of the experiment.
Serval consumes 29.3\% (24.6\%) CPU.
Serval has an additional hashing step, which results in higher CPU load at the start of the experiment.
With only 26.9\% (13.1\%), IBR-DTN is the most efficient tested DTN software in terms of CPU usage.

In terms of network usage, \dtn reaches about 42.0 Mbit/s (19.7 Mbit/s) for transmitting bundles from node to node, while Forban achieves about 32.8 Mbit/s (22.8 Mbit/s).
IBR-DTN and Serval achieve 42.3 Mbit/s (23.7 Mbit/s) and 39.5 Mbit/s (20.0 Mbit/s), respectively.
Although the theoretical total network load for the entire network can be up to 1.674 Gbit/s, the tested DTN software systems used only the maximum bandwidth per link, which is 54 Mbit/s, in peak situations.
This indicates that every DTN software needs to receive the entire bundle before transmitting it to the next node.

To summarize, \dtn requires slightly more CPU utilization than IBR-DTN and Serval, but has the advantage of transmitting files faster than all other DTN systems in most cases, as shown in Section~\ref{subsub:transmission_time}.

%% file: 06_conclusion.tex
\section{Conclusion}
\label{sec:conclusion}
We presented an open source DTN implementation, called \dtn, of the recently released Bundle Protocol BP7 (draft version 13), written in the Go programming language.
\dtn is designed to offer extensibility and supports multiple use cases, such as enabling communication in emergency and disaster scenarios or providing connectivity for rural areas.
Furthermore, we presented results of a comparative experimental evaluation of \dtn and other DTN systems including Serval, IBR-DTN, and Forban.
Our results indicated that \dtn is a flexible and efficient open-source multi-platform implementation of the most recent version of BP7.

There are several areas for future work.
For example, the BP does not define any kind of security or privacy mechanisms, although optional extension exist.
This opens the field of DTN-related security and privacy research based on \dtn.
Furthermore, for sensor networks or deployments in rural areas, \dtn's energy consumption should be evaluated.
Due to \dtn's modular routing interface, new DTN routing algorithms for vehicular ad-hoc networks or UAV-based information dissemination should be investigated.
Finally, new Convergence Layers based on emerging radio technologies, such as LoRa or mmWave communication, could be developed.